\documentclass[conference]{IEEEtran}
\IEEEoverridecommandlockouts
\usepackage{amsmath,amssymb,amsfonts}
\usepackage{algorithmic}
\usepackage{graphicx}
\usepackage{textcomp}
\usepackage{xcolor}
\usepackage{balance}
\usepackage{csquotes}
\usepackage{hyperref}
\usepackage{graphicx}
\usepackage{biblatex}
\usepackage[compact]{titlesec}         
\titlespacing{\section}{0pt}{0pt}{0pt} 
\AtBeginDocument{
  \setlength\abovedisplayskip{0pt}
  \setlength\belowdisplayskip{0pt}}
\usepackage[nolist,nohyperlinks]{acronym}
\begin{acronym}
\acro{HRI}{human-robot interaction}
\acro{HCI}{human-computer interaction}
\acro{HCD}{Human-Centered Design}
\acro{UCD}{User-Centered Design}
\acro{AI}{artificial intelligence}
\acro{NDAs}{non-disclosure agreement}
\acro{EMAR}{Ecological Momentary Assessment Robot}
\end{acronym}

\def\BibTeX{{\rm B\kern-.05em{\sc i\kern-.025em b}\kern-.08em
    T\kern-.1667em\lower.7ex\hbox{E}\kern-.125emX}}
\begin{document}
\title{Participatory Design for Mental Health Data \\ Visualization on a Social Robot\\
}
\vspace{-1mm}
\makeatletter
\newcommand{\linebreakand}{%
  \end{@IEEEauthorhalign}
  \hfill\mbox{}\par
  \mbox{}\hfill\begin{@IEEEauthorhalign}
}
\makeatother
\author{\IEEEauthorblockN{Raida Karim}
\IEEEauthorblockA{\textit{University of Washington}\\
Seattle, Washington, United States\\
rk1997@cs.washington.edu}
\and
\IEEEauthorblockN{Edgar Lopez}
\IEEEauthorblockA{\textit{University of Washington}\\
Seattle, Washington, United States\\
lopeze7@uw.edu}
\linebreakand
\IEEEauthorblockN{Elin A. Björling}
\IEEEauthorblockA{\textit{University of Washington}\\
Seattle, Washington, United States\\
bjorling@uw.edu}
\and
\IEEEauthorblockN{Maya Cakmak}
\IEEEauthorblockA{\textit{University of Washington}\\
Seattle, Washington, United States\\
mcakmak@cs.washington.edu}
}
\maketitle
\begin{abstract}
The intersection of data visualization and human-robot interaction (HRI) is a burgeoning field. Understanding, communicating, and processing different kinds of data for creating versatile visualizations can benefit HRI. Conversely, expressing different kinds of data generated from HRI through effective visualizations can provide interesting insights. Our work adds to the literature of this growing domain. In this paper, we present our exploratory work on visualizing mental health data on a social robot. Particularly, we discuss development of mental health data visualizations using a participatory design (PD) approach. As a first step with mental health data visualization
on a social robot, this work paves the way for relevant further work and using social robots as data visualization tools.  
\end{abstract}
\begin{IEEEkeywords}
Participatory design, mental health, community, data visualization, social robots, human-robot interaction
\end{IEEEkeywords}
\section{Introduction and Background}
Despite many opportunities for collaborative research in data visualization and HRI, not much work has been contributed to this intersection~\cite{szafir}. Some existing works of this area include visualizing data of children's touch patterns on a social robot~\cite{touch}. Contributing to this promising domain's literature, we worked on developing visualizations of mental health data for a social robot. Social robots have been used to support mental health in various ways such as to help children with autism improve on their social skills~\cite{nose}. They have been used to help older adults by reducing feelings of loneliness~\cite{Melinda}, and other populations. However, existing work only shows support for mental health through social robots by responding interactively to human activity to help them learn relevant skills. No work has shown the use of social robots as a means of visualizing mental health data. Therefore, our work is novel or first of its kind. 

\begin{figure*}
\centerline{\includegraphics[width=1.0\textwidth]{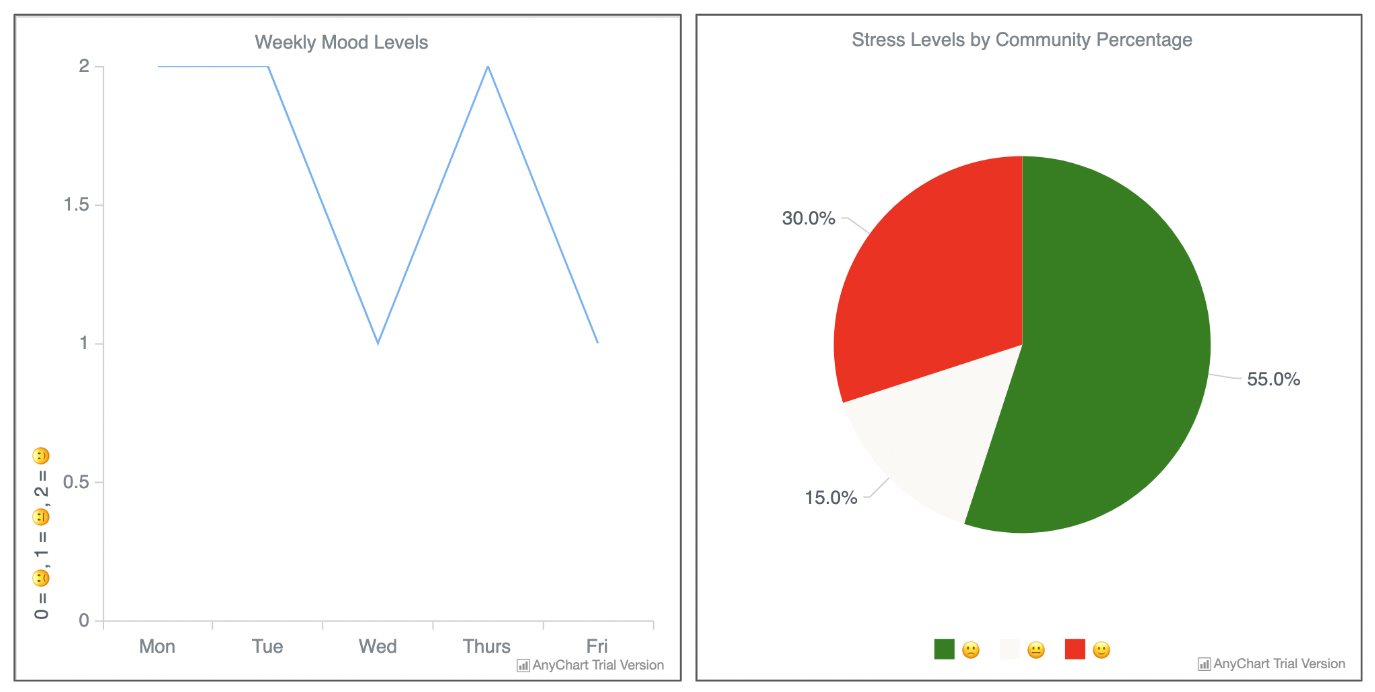}}
\caption{A line chart visualizing mood data (left) and a color-coded pie chart visualizing stress data (right).}
\label{fig:both}
\end{figure*}

\section{A Social Robot \& Mental Health Data}
We detail here the procedure of collecting and visualizing mental health data in these two respective stages:
\subsection{Data Collection}
\vspace{-1mm}
We conducted a five-weekdays HRI study in an American university campus with a total of fifty-five (n=55) participants sharing their in-the-moment mood and stress levels with a social robot. Our previous work~\cite{lbr} showed using an emoji likert scale can enhance coherence and accessibility in portraying different levels of mood or stress data, which is what we used. The users' shared data were stored in a secured firebase \footnote{Firebase: \url{https://firebase.google.com/}}.

\subsection{Data Visualization}
\vspace{-1.8mm}
We developed data visualization software with the updated static data visualization template from~\cite{lbr} in our social robot's software platform. These visualizations are shown in Fig.~\ref{fig:both}, and were implemented in JavaScript using AnyChart library \footnote{AnyChart: \url{https://www.anychart.com/}}. When the visualization program is run on the robot, mood/stress data from firebase is sent to visualization software and data visualizations are created in real-time. 

\section{Discussion}
Mental health data visualizations with a social robot can potentially improve mental health~\cite{lbr}. To the best of our knowledge, this is the first work rendering mental health data visualizations on a social robot seeking to alleviate mental health issues among users. Although this work has been conducted with data collected from people of a university campus, collecting and visualizing other community's data can inform about distinct mental health needs of each community. Moving forward, we plan to expand this work to other community spaces such as high school, and public library. As mentioned earlier, PD method was used to finalize design of data visualizations~\cite{lbr}. PD helped us to get inputs from the community members in designing, developing and refining the features of the robot-rendered data visualizations for mental well-being. We have been using PD in our work for more than two years. We choose to use PD methodology, because PD considers the intended users' and stakeholders' participation throughout the design process and can result in well-informed and usability-tested features of these data visualizations rendered by a social robot. This can eventually help ensure the success of such robotic technologies in supporting mental health. In~\cite{lbr}, we used qualitative analysis method to derive common themes in user feedback in the initial data visualization templates, which informed Fig.~\ref{fig:both} visualizations.

\section{Conclusion \& Future Directions}
In this work, mental health data are visualized through the means of three types of emojis and color codes in the line chart and the pie chart, respectively (see Fig.~\ref{fig:both}). However, other types of mental health data might generate other visualization designs. If a slider scale with a 10-pt likert was used to collect mental health data in a quantitative measure, we might have a scatter plot visualization with plotted numerical data points. More abstract data might be visualized with Parallel Sets~\cite{parallel}. Finding the most effective visualization to express mental health data collected and rendered via social robots remains an open challenge. User studies can be conducted in different community settings to examine human mental models towards these robot-rendered visualizations. Incorporating feedback from the users might lead to improved or future iterations of these visualization tools. This visualization software is open-source \footnote{Visualization software: \url{https://github.com/mayacakmak/emarsoftware}} so it can be adapted and used for other research purposes, such as on robots of various shapes and sizes, and in robot simulations if a physical robot is not accessible. This work marks a first step towards collecting and visualizing mental health data on a social robot at the convergence of HRI and data visualization. In the future, perhaps more varieties of such data visualizations can be produced with the applications of data integration and machine learning.   

\section{Acknowledgements}
This work was funded in part by National Science Foundation: National Robotics Initiative, SES: Award Abstract 1734100 - Design and Development of a Social Robot for Gathering Ecological Momentary Stress Data from Teens.
\printbibliography

@inproceedings{szafir,
author = {Szafir, Daniel and Szafir, Danielle Albers},
title = {Connecting Human-Robot Interaction and Data Visualization},
year = {2021},
isbn = {9781450382892},
publisher = {Association for Computing Machinery},
address = {New York, NY, USA},
booktitle = {Proceedings of the 2021 ACM/IEEE International Conference on Human-Robot Interaction},
pages = {281–292},
numpages = {12},
location = {Boulder, CO, USA},
series = {HRI '21}
}

@article{parallel,
author = {Kosara, Robert and Bendix, Fabian and Hauser, Helwig},
title = {Parallel Sets: Interactive Exploration and Visual Analysis of Categorical Data},
year = {2006},
issue_date = {July 2006},
publisher = {IEEE Educational Activities Department},
address = {USA},
volume = {12},
number = {4},
issn = {1077-2626},
journal = {IEEE Transactions on Visualization and Computer Graphics},
month = {jul},
pages = {558–568},
numpages = {11}
}

@inproceedings{lbr,
author = {Karim, Raida and Zhang, Yufei and Alves-Oliveira, Patr\'{\i}cia and Bj\"{o}rling, Elin A. and Cakmak, Maya},
title = {Community-Based Data Visualization for Mental Well-Being with a Social Robot},
year = {2022},
publisher = {IEEE Press},
booktitle = {Proceedings of the 2022 ACM/IEEE International Conference on Human-Robot Interaction},
pages = {839–843},
numpages = {5},
location = {Sapporo, Hokkaido, Japan},
series = {HRI '22}
}

@article{Melinda,
author = {Kessler, Ronald and Walters, Ellen and Forthofer, Melinda},
year = {1998},
month = {09},
pages = {1092-6},
title = {The Social Consequences of Psychiatric Disorders, III: Probability of Marital Stability},
volume = {155},
journal = {The American journal of psychiatry}
}

@inproceedings{nose,
author = {Costa, Sandra and Lehmann, Hagen and Robins, Ben and Dautenhahn, Kerstin and Soares, Filomena},
year = {2013},
month = {03},
pages = {},
title = {Where is your nose? - Developing body awareness skills among Children with Autism using a humanoid robot}
}

@inproceedings{touch,
author = {Rogers, Kris and Wiles, Janet and Heath, Scott and Sommer, Kristyn and Taufatofua, Jonathon},
year = {2016},
month = {03},
pages = {499-500},
title = {Discovering patterns of touch: A case study for visualization-driven analysis in Human-Robot Interaction}
}
\end{document}